\documentclass[%
	reprint,
superscriptaddress,
 amsmath,amssymb,aps,
 apsrev,
 longbibliography,
]{revtex4-2}

\usepackage{graphicx}
\usepackage{dcolumn}
\usepackage{bm}
\usepackage{units}
\usepackage[colorlinks = true, linkcolor = blue,
            urlcolor  = blue,citecolor = blue,
            anchorcolor = blue]{hyperref}

\usepackage[dvipsnames]{xcolor}
\usepackage{textgreek}
\newcommand{\erf}{\ensuremath{\operatorname{erf}}}

\begin{document}

\preprint{APS/123-QED}

\title{Narrow bandwidth, low-emittance positron beams from a laser-wakefield accelerator}


\newcommand{\JAI}{The John Adams Institute for Accelerator Science, Imperial College London, London, SW7 2AZ, UK}

\newcommand{\GOLP}{GoLP/Instituto de Plasmas e Fus\~{a}o Nuclear, Instituto Superior T\'{e}cnico, U.L., Lisboa 1049-001, Portugal}

\newcommand{\CLF}{Central Laser Facility, STFC Rutherford Appleton Laboratory, Didcot OX11 0QX, UK}

\newcommand{\UCL}{Department of Physics and Astronomy, University College London, London WC1E 6BT, UK}

\newcommand{\LMU}{Fakult\"at f\"ur Physik, Ludwig-Maximilians-Universit\"at M\"unchen, D-85748 Garching, Germany}
\newcommand{\MPQ}{Max-Planck-Institut f\"ur Quantenoptik, Hans-Kopfermann-Str. 1, D-85748 Garching, Germany}

\newcommand{\DESY}{Deutsches Elektronen-Synchrotron DESY, Notkestr. 85, 22607 Hamburg, Germany}

\newcommand{\CI}{The Cockcroft Institute, Keckwick Lane, Daresbury, WA4 4AD, United Kingdom}

\newcommand{\LANCS}{Physics Department, Lancaster University, Lancaster LA1 4YB, United Kingdom}

\newcommand{\UMICH}{Center for Ultrafast Optical Science, University of Michigan, Ann Arbor, MI 48109-2099, USA}

\newcommand{\SUPA}{SUPA, Department of Physics, University of Strathclyde, Glasgow G4 0NG, UK}

\newcommand{\LLNL}{Lawrence Livermore National Laboratory (LLNL), P.O. Box 808, Livermore, California 94550, USA}

\newcommand{\DLS}{Diamond Light Source, Harwell Science and Innovation Campus, Fermi Avenue, Didcot OX11 0DE, UK}

\newcommand{\YORK}{York Plasma Institute, Department of Physics, University of York, York YO10 5DD, UK}

\newcommand{\ELI}{ELI-Beamlines, Institute of Physics, Academy of Sciences of the Czech Republic, 18221 Prague, Czech Republic}

\newcommand{\LUND}{Department of Physics, Lund University, P.O. Box 118, S-22100, Lund, Sweden}

\newcommand{\QUB}{School of Mathematics and Physics,
  Queen's University Belfast
 , BT7 1NN, Belfast UK}

\author{M.J.V.~Streeter}
\affiliation{\QUB}

\author{C.~Colgan}
\affiliation{\JAI}

\author{J.~Carderelli}
\affiliation{\UMICH}

\author{Y.~Ma}
\affiliation{\UMICH}

\author{N.~Cavanagh}
\affiliation{\QUB}

\author{E.E.~Los}
\affiliation{\JAI}

\author{H.~Ahmed}
\affiliation{\CLF}

\author{A.F.~Antoine}
\affiliation{\UMICH}

\author{T.~Audet}
\affiliation{\QUB}

\author{M.D.~Balcazar}
\affiliation{\UMICH}

\author{L.~Calvin}
\affiliation{\QUB}

\author{B.~Kettle}
\affiliation{\JAI}

\author{S.P.D.~Mangles}
\affiliation{\JAI}

\author{Z.~Najmudin}
\affiliation{\JAI}

\author{P.P.~Rajeev}
\affiliation{\CLF}

\author{D.R.~Symes}
\affiliation{\CLF}

\author{A.G.R.~Thomas}
\affiliation{\UMICH}

\author{G.Sarri}
\email{g.sarri@qub.ac.uk}
\affiliation{\QUB}

\date{\today}

\begin{abstract}
The rapid progress that plasma wakefield accelerators are experiencing is now posing the question as to whether they could be included in the design of the next generation of high-energy electron-positron colliders. 
However, the typical structure of the accelerating wakefields presents challenging complications for positron acceleration. 
Research in plasma-based acceleration of positrons has thus far experienced limited experimental progress due to the lack of positron beams suitable to seed a plasma accelerator. 
Here, we report on the first experimental demonstration of a laser-driven source of ultra-relativistic positrons with sufficient spectral and spatial quality to be injected in a plasma accelerator.
Our results indicate, in agreement with numerical simulations, selection and transport of positron beamlets containing $N_{e+}\geq10^5$ positrons in a 5\% bandwidth around 600 MeV, with femtosecond-scale duration and micron-scale normalised emittance. 
Particle-in-cell simulations show that positron beams of this kind can be efficiently guided and accelerated in a laser-driven plasma accelerator, with favourable scalings to further increase overall charge and energy using PW-scale lasers. 
The results presented here demonstrate the possibility of performing experimental studies of positron acceleration in a plasma wakefield. 
\end{abstract}

\maketitle
Plasma-based wakefield accelerators \cite{Esarey2009RMP} have been gathering significant attention in recent years, mainly thanks to the ultra-high accelerating gradients (in the region of \unit[10s - 100]{GV/m})  that they are able to sustain, providing a promising platform for the miniaturisation of particle accelerators. In addition, electron beams from a plasma accelerator naturally possess unique properties such as intrinsic femtosecond-scale duration, micron-scale source size, and sub-micron normalised emittances at the GeV level (see, e.g., Ref. \cite{Weingartner2012PRAB,Lundh2011NP}). Progress in laser and plasma technology can now also enable stable operation of these accelerators over long periods \cite{Maier2020PRX}. 

This rapid scientific and technological progress is now posing the realistic question as to whether plasma-based acceleration could be a viable complementary technology for the next generation of particle colliders, proposed to break the TeV barrier \cite{Adli2019PTRC}. Several international consortia and proposed large-scale facilities are now actively addressing this question (e.g., \cite{EuPRAXIA2020EPJ,ALEGRO2019}), which is also identified by several national and international roadmaps as a central area of research (see, e.g., Refs. \cite{PWASC2019arxiv,USAroadmap2016,Adolphsen2022arxiv}). 

While plasma-based acceleration of electrons has achieved a relatively high level of maturity, plasma-based acceleration of positrons presents harder fundamental challenges. 
This is due to the inherent structure of the wakefield accelerating structures \cite{Lotov2007POP} as well  as the stringent requirements on the temporal and spatial properties of the seed positron beam. 
For example, quasi-linear acceleration in a plasma with a density of \unit[$n_e=10^{17}$]{cm$^{-3}$} (as proposed in baseline stuides, see e.g., Ref.\cite{Schroeder2010PRSTAB}), would require longitudinal and transverse seed beam dimensions \unit[$\sigma_z, \sigma_x \lesssim 10$]{{\textmu}m}.  
While several positron acceleration schemes have been theoretically proposed \cite{Kimura2011PRAB,Yu2014POP,Vieira2014PRL,Yi2014SR,Jain2015PRL,Diederichs2019PRAB,Silva2021PRL} and first proof-of-principle experiments have demonstrated potential in this direction \cite{Muggli2008PRL,Corde2015N,Gessner2016NC,Doche2017SR,Lindstrom2018PRL}, progress in this area has been hampered by the scarcity of facilities capable of providing positron beams with these demanding characteristics. To date, only FACET-II\cite{Yakimenko2019PRAB} at SLAC could be in principle suited in the future for proof-of-principle experiments in this area.

In order to enable experimental studies of positron acceleration in a wakefield, it is thus necessary first to provide positron witness beams with high spatial and spectral quality at the GeV level. To achieve this goal, it would be desirable to avoid storage rings, so that the intrinsic femtosecond-scale duration of laser-driven positrons could be preserved. Laser-driven generation of ultra-relativistic positron beams is thus currently being actively studied, with several landmark results already reported, including maximum positron energies in the region of \unit[100]{MeV} \cite{Sarri2013PRL,Alejo2020PPCF,Li2019SA}, the generation of high-density and quasi-neutral electron-positron beams \cite{Sarri2015NC}, and first experimental observation of pair-plasma dynamics \cite{Warwick2017PRL}. 
However, to effectively enable plasma acceleration of positrons in the laboratory, it is necessary to produce beams that simultaneously have \%-level energy spreads, femtosecond-scale duration, and micron-scale normalised emittance \cite{Adli2019PTRC,ALEGRO2019}. 
To date, properties of this kind have only been predicted numerically \cite{Alejo2019SR,Alejo2020PPCF,Sarri2022PPCF}.

\begin{figure*}[t!]
  \centering
  \includegraphics[width=17.9cm]{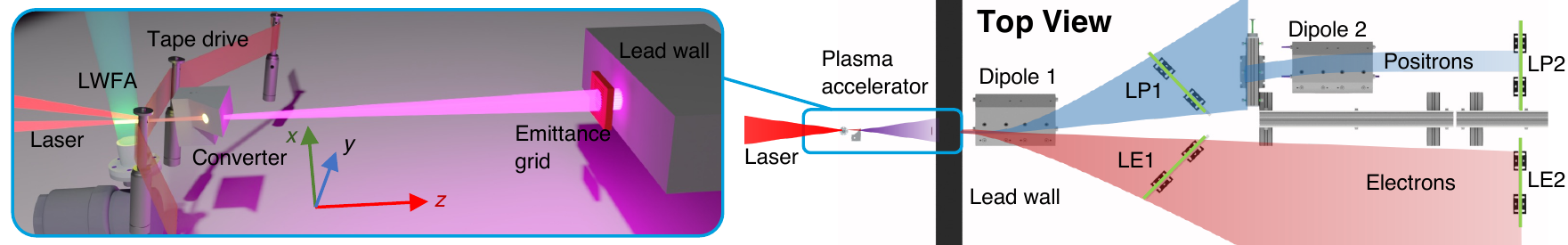} 
  \caption{
\textbf{Illustration of the experimental setup} showing the electron plasma accelerator, the converter, the emittance mask, scintillators for electrons (LE1 and LE2) and positrons (LP1 and LP2). Electron (red) and positron (blue) trajectories are also shown to guide the eye.
}
  \label{fig:exp_setup}
\end{figure*}

Here, we experimentally demonstrate that  positron beams of this kind can be generated with a laser system of relatively modest peak power. 
Crucially, we demonstrate that the obtained positron beams are of sufficient quality to be energy-selected, with our results showing the isolation of positron beamlets containing $\geq$ $10^5$ positrons in a 5\% bandwidth at energies exceeding 500 MeV. These beams present femtosecond-scale duration and micron-scale normalised emittance, and are thus of sufficient spectral and spatial quality to efficiently act as a witness beam in a positron wakefield accelerator, as we demonstrate with proof-of-principle particle-in-cell simulations. 
Favourable scalings in the beam spectral and spatial characteristics already indicate that even higher quality can be achieved with laser systems of higher peak power \cite{Alejo2019SR,Sarri2022PPCF}.
These results represent a critical milestone towards the realisation of plasma-based particle accelerators and their potential implementation in the next generation of particle colliders.

\section*{Results}
\vspace{-10pt}
\noindent\textbf{Experimental setup.}
The experiment was performed using the Gemini laser at the Central Laser Facility \cite{Hooker2006JDP}, (setup sketched in figure \ref{fig:exp_setup}).
The laser pulses contained a mean (and RMS variation) of \unit[$7.9\pm0.5$]{J} in a FWHM pulse length of \unit[$48\pm7$]{fs} (peak power \unit[$P_0 = 156\pm9$]{TW}) and a central wavelength of \unit[800]{nm}.
The pulses were focused with an $f/40$ off-axis parabola into a gas jet from a \unit[15]{mm} exit diameter nozzle, to generate high-energy electron beams via laser wakefield acceleration (LWFA).
The gas was a mixture of 2\% nitrogen and 98\% helium and had an electron density of \unit[$n_e = (1.5\pm0.2)\times10^{18}$]{cm$^{-3}$}, as measured via optical interferometry.
The laser focal spot, measured in vacuum, was \unit[$(37\pm 3)\times (52\pm 4)$]{{\textmu}m} in the transverse $x$ and $y$ directions respectively ($1/e^2$ radius), giving a peak intensity \unit[$I_0 = (3.1 \pm 0.3) \times 10^{18}$]{Wcm$^{-2}$.}

The residual laser exiting the LWFA was removed by reflection from a self-generated plasma mirror on the surface of a \unit[125]{{\textmu}m} polyimide tape which was replenished after every shot. 
The tape target was kept for all the experimental data shown here.
A movable lead converter target (placed at a distance $z_D$ = \unit[50]{mm} from the LWFA exit plane of the gas jet) was used to generate electron-positron beams through a two-step bremsstrahlung induced Bethe-Heitler process \cite{Sarri2013PPCF}.
The converter was a 45-degree wedge, such that translating it perpendicularly to the electron beam axis allowed the effective converter thickness to be varied continuously over the range \unit[$1\geq L \geq 25$]{mm}.

A shielding lead wall with an on-axis \unit[10]{mm} diameter aperture was placed to allow only particles emitted from the converter within a \unit[12.6]{mrad} half-angle to propagate to the detectors.
A permanent magnetic dipole (Dipole 1 in figure \ref{fig:exp_setup}, with integrated strength of \unit[$B_x z = 0.3$]{Tm}) was placed behind the lead wall to sweep electrons and positrons onto the primary scintillator screens (LP1 and LE1, both Kodak LANEX) either side of the central axis,
allowing observation of particles with kinetic energy \unit[$E\geq200$]{MeV}. 
Due to experimental limitations, the two screens were placed at slightly different distances from the dipole and angles from the main axis: 134 cm and 41.1$^\circ$ for the positron side of the spectrometer (LP1 in Fig. \ref{fig:exp_setup}) and 137 cm and 45.4$^\circ$ for the electron side of the spectrometer (LE1 in Fig. \ref{fig:exp_setup}).
A second scintillator screen (LE2) was placed \unit[$1$]{m} behind the first in order to increase the measurement accuracy of the high energy electrons (more details in the Methods section). A second identical magnetic dipole (Dipole 2 in figure \ref{fig:exp_setup}) with a \unit[$25$]{mm} wide lead slit placed at its entrance was positioned in the dispersed positron beam, in a dog-leg configuration.
The slit performed energy selection on the dispersed positron beam, which was then collimated onto an additional scintillator screen (LP2) by the magnet. 

 A \unit[5.0]{mm} thick tungsten mask composed of horizontal slits with a period of \unit[1100]{{\textmu}m} (\unit[550]{{\textmu}m} gaps) was placed into the beam axis \unit[290]{mm} behind the rear face of the converter (see figure \ref{fig:exp_setup}) to perform energy-resolved emittance measurements on the generated positrons and the scattered electrons (details on the emittance retrieval in the Methods section).

\noindent\textbf{Experimental results.}
\begin{figure}[t!]
  \centering
  \includegraphics[width=8.5cm]{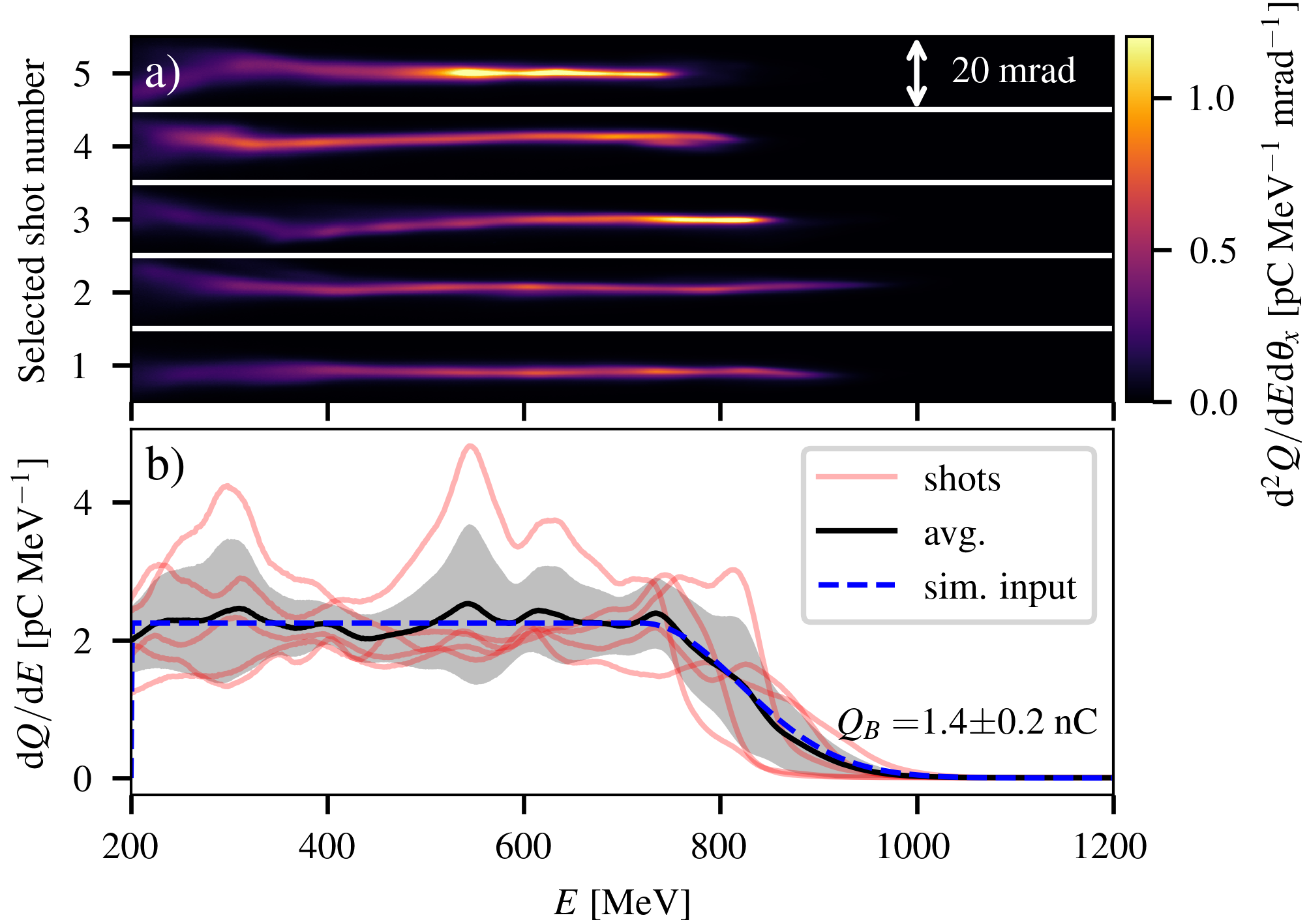} 
  \caption{\textbf{Primary electron beam characteristics.}
  Typical a) angularly resolved and b) angularly integrated electron spectra of the LWFA electron beams (red).
  The average (black) and standard deviation (grey) of the integrated electron spectra are shown along with the approximation used as an input for the positron generation simulations (blue).
}
  \label{fig:input_spectrum}
\end{figure}
The electron spectra produced by the LWFA were first characterised with the tape drive in place but without the converter.
Ten shots were taken with nominally identical conditions, with variations in the electron spectrum due to inherent variations in laser and plasma source parameters.
The five shots with the highest total beam energy are shown in figure \ref{fig:input_spectrum}a.
The angularly integrated electron spectra for each of these shots are plotted in figure \ref{fig:input_spectrum}b along with their average.
When analysing the electron-positron beams, shots with the highest total charge (3-8 shots out of 10) were used for each converter length, and so the average shown in figure \ref{fig:input_spectrum}b was taken as the expected LWFA spectrum for those shots.
The average (and RMS variation) total beam charge for electron energies above \unit[200]{MeV} was \unit[$Q_b=1.4\pm0.2$]{nC} and the total beam energy was \unit[$W_b =0.8\pm0.1$]{J}, giving a laser-to-electron beam energy efficiency of $\eta \approx 10 \%$.
The angular distribution of the energy integrated electron spectra is partially affected by the propagation through the tape target \cite{Raj2020PRR} and is closely approximated by the square of a Lorentzian function with a FWHM of \unit[$\theta_x = (3.8\pm0.4)$]{mrad}.

\begin{figure}[b!]
   \centering
   \includegraphics[width=8.5cm]{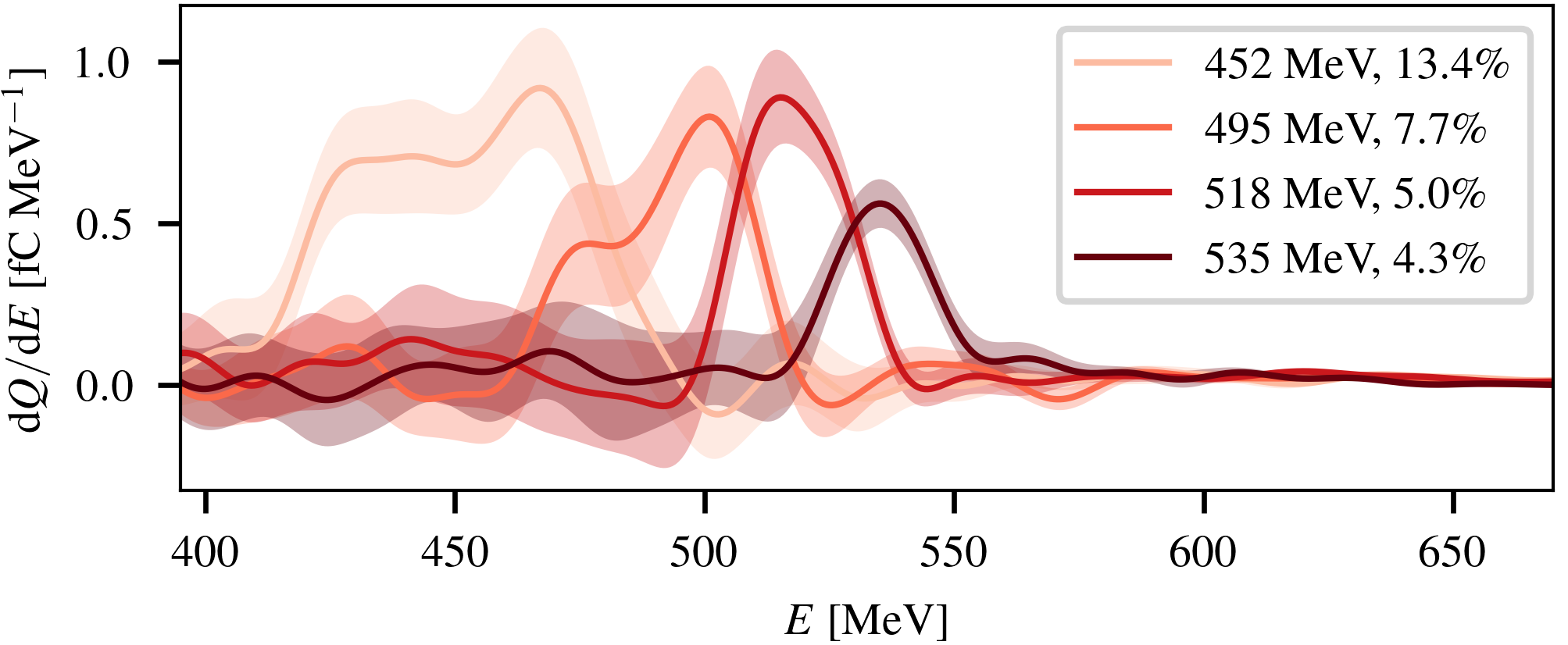}
   \caption{
   \textbf{Narrow energy spread positron beams.} Typical single-shot positron spectra measured after energy selection for different positions of the energy selection slit.
   Raw data has been background-subtracted and smoothed with a \unit[10]{MeV} Gaussian filter, with the shaded region representing the local RMS scatter of the data.
   The central energy and FWHM bandwidth of each spectrum is indicated in the figure legend.
    }
   \label{fig:EnergySelection}
\end{figure}

To demonstrate energy selection of laser-driven positron beams, a lead converter thickness of  \unit[$5.0$]{mm} ($0.9$ radiation lengths) was used and the recorded positron spectra after the second dipole magnet are shown in figure \ref{fig:EnergySelection} for different transverse positions of the slit. For central energies of \unit[$E>500$]{MeV}, more than $10^5$ positrons per shot were transmitted within a FWHM bandwidth of $\Delta E/E \leq5\%$, demonstrating the possibility of performing efficient energy selection and capture of laser-driven positron beams of this kind. 
Similar results were obtained for different target thicknesses (not shown).

\begin{figure}[b!]
   \centering
   \includegraphics[width=8.5cm]{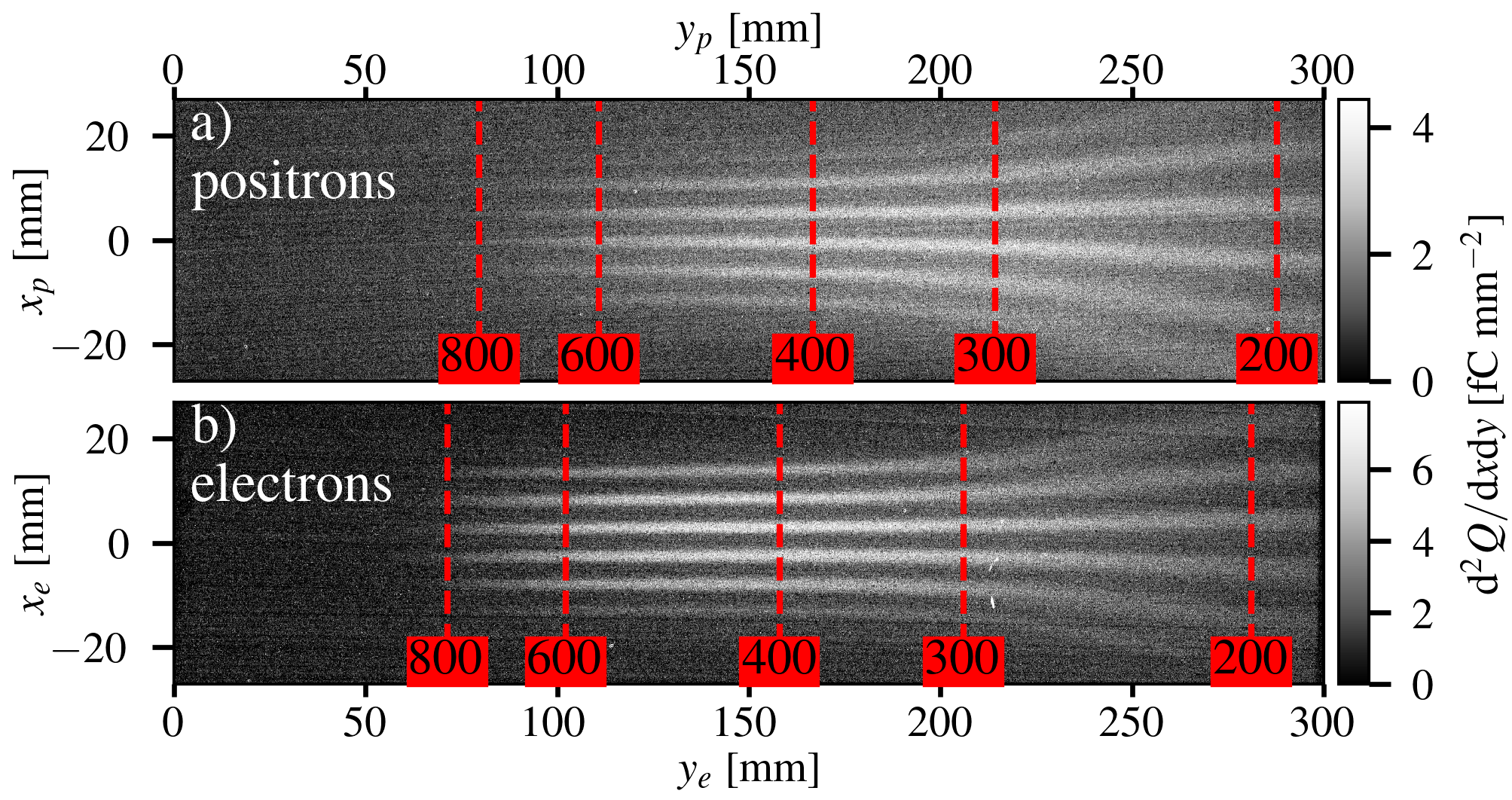}
   \caption{
   \textbf{Raw images of energy-resolved beam profiles with the emittance mask.} Example modulated a) positron and b) electron spatial charge density as a function of position on the screens ($x_p, y_p, x_e, x_e$) for a single shot with a converter thickness of \unit[8.0]{mm} and the emittance mask in the beam-line.
   The positions corresponding to the given particle energies in MeV are shown as vertical red dashed lines.
   The slight difference between the electron and positron raw data is due to the slightly different position of the scintillator screens (discussed in the text).
    }
   \label{fig:exampleGridSpectra}
\end{figure}

The energy-resolved emittance of the electrons and positrons exiting the converter target  were characterised for different converter thicknesses by inserting the tungsten mask in front of the lead sheilding wall (see Fig. \ref{fig:exp_setup}).  
This insertion resulted in a modulated beam profile onto the scintillators, as shown in figure \ref{fig:exampleGridSpectra}.
The modulation had a lower spatial frequency at low energy, since 
particles with an energy \unit[$E\lesssim300$]{MeV} exited the side of the dipole field and experienced 
defocusing as their longitudinal momentum ($p_z$) coupled with the transverse fringe field ($B_y$).
In order to correct for the fringe field defocusing effect in the analysis, the measured signals were re-scaled in the non-dispersion direction such that the magnification of the grid pattern was kept constant for all energies. 
This results in an overestimation of the source size by $\sim 5\%$ for energies \unit[$E<300$]{MeV}. 
In addition, the finite spectral resolution of the spectrometer caused blurring of the grid pattern where the magnification varied most strongly.
As a result, the source size has a total systematic uncertainty of $\sim30$\% for \unit[$E<350$]{MeV}. 
For \unit[$E>350$]{MeV}, the blurring effect had a negligible effect on the measured source size.

\begin{figure*}[hbpt]
   \centering
   \includegraphics[width=18cm]{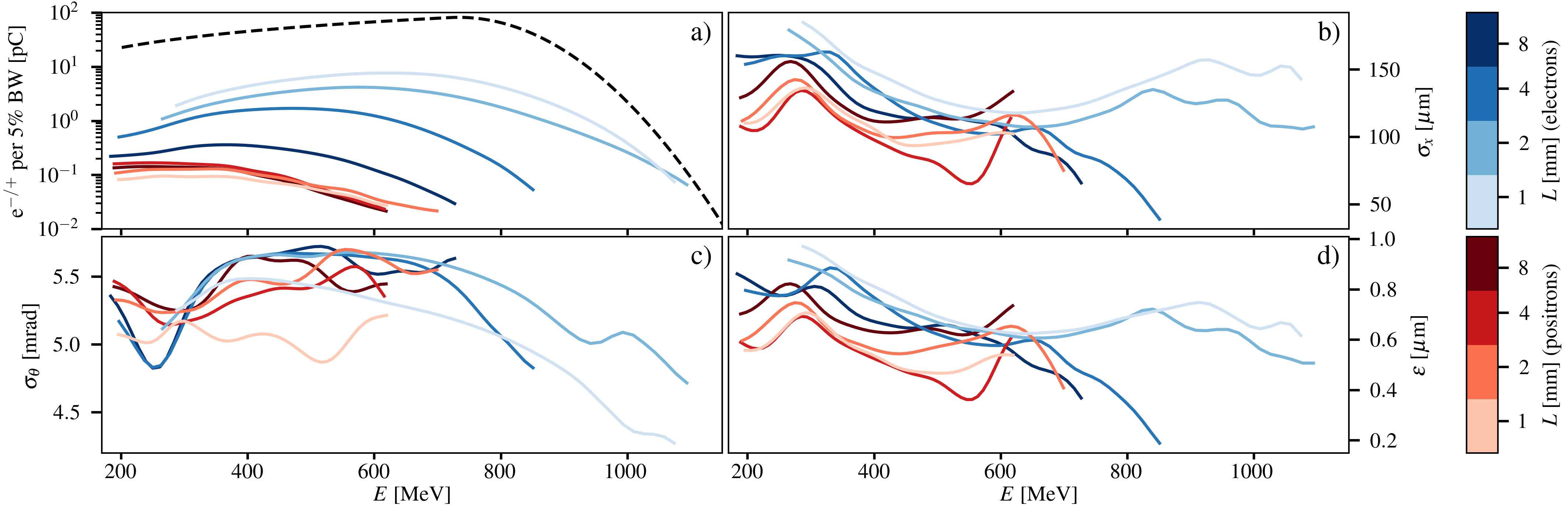}
   \caption{
   \textbf{Positron properties as function of energy and converter thickness.} Measured electron and positron beam properties as functions of particle energy for different converter thicknesses.
   The a) spectra (charge per 5\% bandwidth), b) source size, c) divergence and d) geometric emittance are given for each converter length as shown by the color-bars at the side of the figure.
   For each converter thickness, only the shots resulting in the highest charge of the positron beams were used for the analysis; the lines shown are thus an average of 4, 6, 3 and 8 shots for converter lengths of \unit[1.0, 2.0, 4.0 and 8.0]{mm}, respectively (RMS variation of $\approx$ 43\% in the spectrum and $\approx$ 20\% in emittance, respectively).
   Each line results from a Gaussian weighting of each measurement point using a kernel width of \unit[$\sigma_E=25$]{MeV}.
   The typical input electron spectrum (black-dashed) is shown in a) for comparison.
    }
   \label{fig:ExperimentalBeamProperties}
\end{figure*}

The beam divergence $\sigma_{\theta} = \sqrt{\langle x'^2\rangle}$ was found for each energy slice by fitting the envelope of the background subtracted signal and dividing by the source-to-screen distance.
The non-zero source size (assumed to be Gaussian \cite{Alejo2019SR}) of the electron and positron beams resulted in a convolution of the detector resolution limited grid pattern with a Gaussian distribution with an RMS width of $\sigma_s = \sigma_x M$.
The value $\sigma_s$ was found by iterative deconvolution, and then divided by the source plane magnification, $M$, to yield the source size (RMS) $\sigma_x = \sqrt{\langle x^2\rangle}$.

The geometric emittance of a particle beam is defined as $\epsilon = \sqrt{\langle x^2\rangle \langle x'^2 \rangle - \langle x x'\rangle }$, where $\langle x x'\rangle$ is the angle-position correlation term.
In the drift space between the LWFA and the converter, the primary electron beam develops a strong correlation term. 
However, the relatively large scattering angles in the converter dominate so that, at the exit of the converter, the positron beam is largely uncorrelated. 
Monte-Carlo simulations (discussed in the following) indicate that the small level of remaining correlation implies that the positron beam is equivalent to an uncorrelated beam originating from \unit[$100-200$]{{\textmu}m} inside the rear surface of the converter.
Therefore, the correlation term was neglected and the geometric emittance was calculated as the product of the measured divergence and source size, i.e. $\epsilon = \sigma_x \sigma_{\theta}$.

The electron and positron beam properties were measured as functions of energy for converter lengths \unit[$L=1.0, 2.0, 4.0, 8.0$]{mm} (0.2, 0.4, 0.7 and 1.4 radiation lengths) and are plotted in figure \ref{fig:ExperimentalBeamProperties}. 
The number of observed electrons was seen to decrease with converter thickness (figure \ref{fig:ExperimentalBeamProperties}a), while the number of positrons was maximised for \unit[$L=4.0$]{mm}.
The electron and positron source size (figure \ref{fig:ExperimentalBeamProperties}b) was observed to weakly decrease with energy, from \unit[140$\pm$10]{{\textmu}m} at \unit[300]{MeV} down to \unit[110$\pm$10]{{\textmu}m} at \unit[600]{MeV}, with fluctuations between 200 and \unit[300]{MeV} due to the systematic uncertainties described before.
The electron and positron RMS divergence (figure \ref{fig:ExperimentalBeamProperties}c) was approximately constant at \unit[$\sigma_{\theta} = 5.3 \pm 0.3$]{mrad}, as it was limited by the aperture in the lead wall. 
Due to the fixed beam divergence, the emittance trends were largely determined by the variation in source size. The positron geometric emittance exhibited a gradual linear decrease as a function of energy with values of \unit[$\epsilon=640$]{nm} at \unit[$E=200$]{MeV} and \unit[$\epsilon=480$]{nm} at \unit[$E=600$]{MeV} for a \unit[1.0]{mm} converter. 
As numerically predicted previously \cite{Alejo2020PPCF}, the positron geometric emittance (figure \ref{fig:ExperimentalBeamProperties}d) was consistently lower than that of the scattered electrons.
The results had an RMS variation of $\approx 43\%$ in the spectrum and $\approx20\%$ in emittance, due to the shot-to-shot variation in the primary electron beam.


In order to ascertain the effect of the beam-line geometry on the measured positron beam characteristics, simulations were performed using the particle physics Monte-Carlo code FLUKA \cite{FLUKA2005,FLUKA2014} (details in the Methods section).
Electrons were initialised from the LWFA electron spectrum approximation shown in figure \ref{fig:input_spectrum}b and the momenta and position of all electrons and positrons were recorded as they exited the rear surface of the converter. 
The divergence of the primary LWFA electron beam was modelled by applying randomised shifts (matching the measured LWFA divergence) to the position and propagation angles of each generated particle.
The transverse particle positions were also modified according to the expected LWFA electron source size of \unit[1]{{\textmu}m} \cite{Kneip2010NP,Corde2011PRL,Wenz2015NC}, although this contribution was observed to be negligible.
The effect of the aperture in the beamline was simulated by removing all particles which had radial positions greater than \unit[5]{mm} at the aperture plane.

\begin{figure}[b!]
   \centering
   \includegraphics[width=8.5cm]{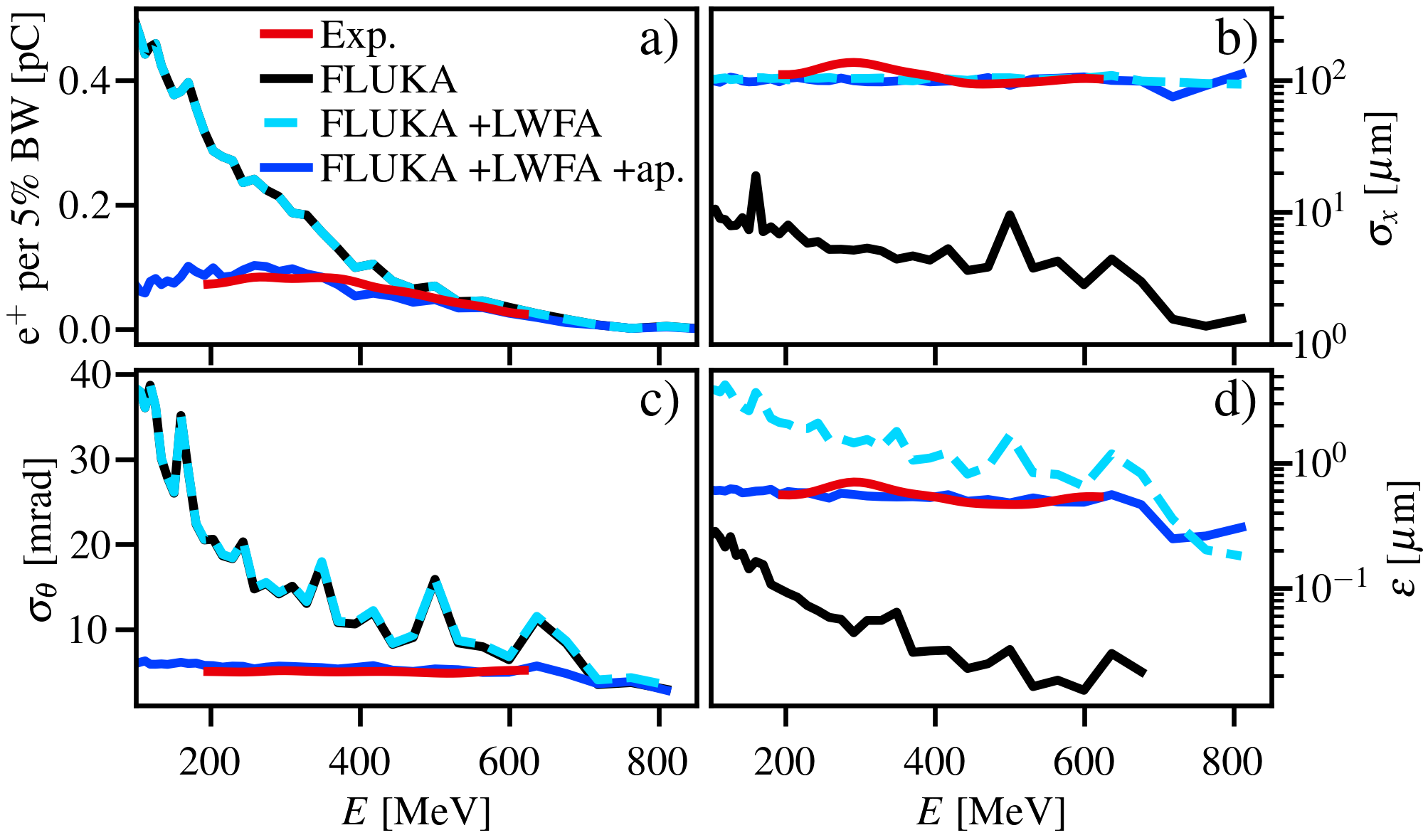}
   \caption{
   \textbf{Comparison with numerical simulations.} Positron beam a) spectrum (charge per 5\% bandwidth), b) source size, c) divergence and d) geometric emittance plotted for a \unit[1.0]{mm} thick lead converter.
   The experimental data (red) is plotted alongside FLUKA simulations for: zero drift distance for the primary electron beam (black); including the drift distance and the  primary electron beam divergence and source size (cyan dashed); and including the \unit[12.6]{mrad} shielding aperture (blue).
   }
   \label{fig:PositronsSourceAperture}
\end{figure}

For example, the results of the numerical simulation for a converter thickness of \unit[$L=1.0$]{mm} show good agreement with the experimental data (blue and red lines in figure  \ref{fig:PositronsSourceAperture}, respectively) with an RMS average difference for all converter lengths of 15\%, 3.5\%, 2.8\% and 0.5\% for the spectrum, emittance, source size and divergence, respectively.

Due to the higher initial divergence for lower energy particles, the aperture transmits fewer lower energy positrons, modifying the detectable spectrum (figure \ref{fig:PositronsSourceAperture}a). 
The aperture also constrains the beam divergence to an approximately constant value of \unit[$\sigma_{\theta}\approx 5$]{mrad} (figure \ref{fig:PositronsSourceAperture}c). 
Including the finite divergence of the LWFA does not affect the positron spectrum or divergence but has a strong effect on the source size (figure \ref{fig:PositronsSourceAperture}b) and, subsequently, the emittance of the positron beam (figure \ref{fig:PositronsSourceAperture}d).
A significant reduction in positron source size can be readily obtained by placing the converter closer to the exit of the LWFA, immediately after the tape drive or, alternatively, by using a beam transport system to minimise the electron beam size on the converter.
A replenishing tape (such as the one used for this experiment) can be used to extract the post-plasma laser pulse and protect the converter from damage. 
Operations of tapes of this kind up to the kHz has already been demonstrated \cite{Haney1993AO}.
As observed in our experiment, the laser-plasma interaction at the plasma mirror surface causes a small (\unit[$\sim$]{mrad}) increase in electron beam divergence \cite{Raj2020PRR}, but this would be negligible compared to the inherent divergence of the pair production process.

For the measured LWFA electron beam FWHM divergence of \unit[3.8]{mrad}, a typical beam waist at the LWFA source of \unit[1]{{\textmu}m}, and a converter thickness of 1 mm, the positron source size at 600 MeV could be reduced to \unit[2.7]{{\textmu}m} by minimising the free drift distance of the electron beam.  
In this case the positron beam would have a divergence $\sigma_\theta=$\unit[5.5]{mrad}, a geometric emittance $\epsilon=$\unit[15]{nm}, and a normalised emittance \unit[$\bar{\epsilon} = \gamma \beta_z \epsilon = 18$]{{\textmu}m} at \unit[600]{MeV}. 
All positron beam parameters for different target thicknesses can then be readily obtained using known scaling laws (see, e.g., Refs. \cite{Alejo2019SR,Sarri2022PPCF}). 
For a converter target placed right after the LWFA, our simulations indicate that different electron source size or divergence will have a small effect on the positron beam characteristics, which can anyway be taken into account by adding it in quadrature to the obtained values. 
For example, doubling the electron source size only increases the positron source size by approximately 10\%.
The simulations also show a spreading in the duration of the positron beam of \unit[0.1]{fs} for positrons within a 5\% bandwidth of \unit[600]{MeV} implying that the positron beam duration will be similar to that of the primary electron beam (i.e., \unit[$\sigma_z \lesssim \lambda_p/2 = 14$]{{\textmu}m} for \unit[$n_e = 1.5 \times 10^{18}$]{cm$^{-3}$}, corresponding to \unit[$\tau\lesssim 50$]{fs}).

\section*{Positron Post-Acceleration Simulations}

Simulations were performed using the particle-in-cell code FBPIC \cite{Lehe2016CPC} to assess the suitability of the generated positron beams for post-acceleration in a plasma wakefield. 
As an example following a baseline study on plasma accelerators \cite{Schroeder2010PRSTAB}, the laser-wakefields were produced by the interaction of a Gaussian laser pulse with radius \unit[$r_L = 70$]{{\textmu}m}, pulse length \unit[$\tau_L = 56$]{fs} and a normalised vector potential of $a_0=1.5$, with a pure helium plasma with an electron density of \unit[$2\times10^{17}$]{cm$^{-3}$} and a length of 10 cm enclosed by two 0.5mm-long linear ramps.

The initial positron particle distribution was obtained from FLUKA simulations of the interaction of the LWFA electron beam with a \unit[1]{mm} converter. 
A longitudinal spread of \unit[10]{{\textmu}m} (uniform distribution) was added to the particles to take the positron bunch duration into account. 
The positron bunch was initialised at the peak of the positive accelerating field of the plasma wakefield. 
Further details of the simulation setup are given in the Methods section.

\begin{figure}[t!]
   \centering
   \includegraphics[width=8.5cm]{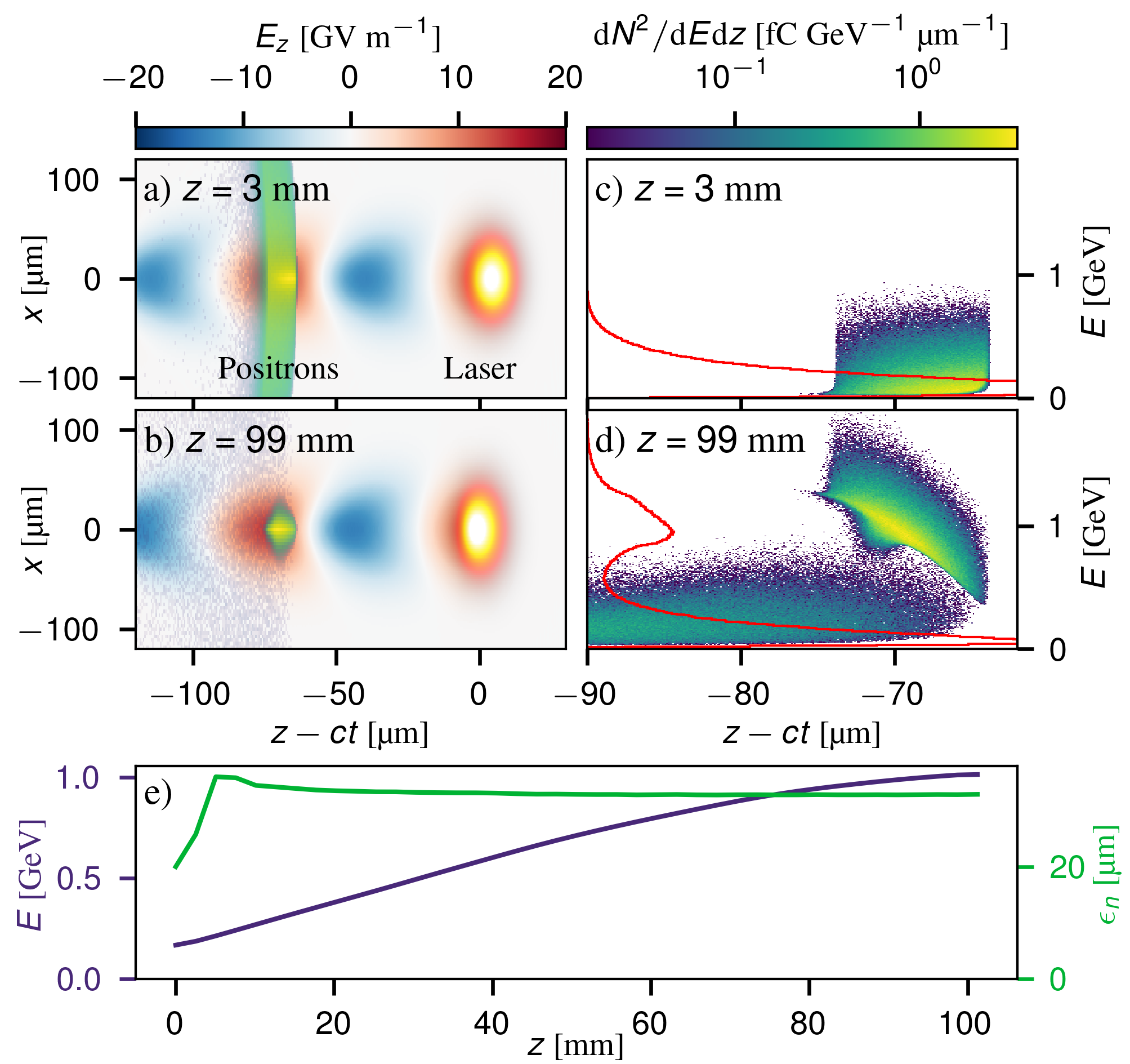}
   \caption{\textbf{Simulated post-acceleration of a laser-generated positron beam.}
   \textbf{a}\&\textbf{b} show the longitudinal electric fields of the plasma wakefield generated by the laser pulse (yellow orb) and the trailing positron bunch (density in a logarithmic colorscale) at the beginning of the plasma and after \unit[96]{mm} of propagation.
   Panels \textbf{c}\&\textbf{d} show the positron longitudinal phase space before an after acceleration, with the energy spectra indicated by the red lines.
   \textbf{e} shows the average energy and normalised emittance of the trapped bunch, defined as being comprised of particles which remain within \unit[$\pm50$]{{\textmu}m} of the central axis.}
   \label{fig:PositronsAccelResults}
\end{figure}

Fig. \ref{fig:PositronsAccelResults} shows a summary of the main results of the simulation. 
For these laser and plasma conditions, a quasi-linear wakefield is generated behind the laser pulse (Fig. \ref{fig:PositronsAccelResults}a) with a peak accelerating field of \unit[13]{GVm$^{-1}$} and a period of \unit[75]{{\textmu}m}. 
After \unit[96]{mm} of propagation in the plasma, 31\% of the charge remains in a focused positron beam (figure \ref{fig:PositronsAccelResults}b) which is accelerated to an average of \unit[1.0]{GeV} with a relative rms energy spread of 20\% (\ref{fig:PositronsAccelResults}c-d).
The beam is chirped, however,  and so the energy spread could be reduced further using a plasma dechirper \cite{Darcy2019PRL}.
After an initial increase, the normalised emittance of the trapped positron beam remains approximately constant throughout the acceleration, at \unit[$\bar{\epsilon}=33.0\pm0.2$]{{\textmu}m}.

These results can in principle be experimentally achieved by driving a laser wakefield accelerator directly behind the converter, as could be achieved with compact plasma mirror staging. This configuration can be directly implemented in existing and near-term laser facilities, enabling experimental studies of wakefield acceleration of positrons.
Alternatively, a magnetic beam-line, as considered for example in EUPRAXIA \cite{EuPRAXIA2020EPJ}, could be used to perform energy selection and controlled focusing to increase the experimental capabilities. As an example of this capability, we show as supplementary material FBPIC simulations of wakefield acceleration of the energy-selected positron bunch (energy of 500 MeV with a 5\% energy spread) in the same conditions as those discussed above \cite{suppl}. 
In this case, approximately 50\% of the positron bunch is trapped and accelerated up to an energy of \unit[$1.2\pm0.3$]{GeV} with a normalised emittance of \unit[57]{{\textmu}m}.  


\section*{Discussion and conclusions}
With the minimisation of the drift length for the LWFA electrons, numerical simulations show that the obtained positron beam characteristics are now well suited for efficient capture and post-acceleration as a witness beam in a plasma wakefield accelerator. 
This result is also consistent with other studies reported in the literature. For example, Silva \emph{et al.} \cite{Silva2021PRL} numerically demonstrated plasma acceleration for a positron bunch with a radius of \unit[$\sigma_x = 5$]{{\textmu}m} and a normalised emittance of \unit[10]{{\textmu}m} using an electron beam driver, while Vieira \emph{et al.} \cite{Vieira2014PRL} numerically demonstrated effective wakefield acceleration of a \unit[$\sigma_x = 6$]{{\textmu}m} positron beam in a Laguerre-Gaussian laser mode. 
Proof-of-principle experiments on wakefield acceleration of positrons reported in the literature \cite{Muggli2008PRL,Corde2015N,Gessner2016NC,Doche2017SR}, used witness positron beams with similar geometric emittance to our inferred beam properties, though at a higher energy (see Table \ref{tbl:beamComparison}), demonstrating the suitability of the positron source reported here to provide witness beams for wakefield acceleration studies, such as proposed for EuPRAXIA \cite{EuPRAXIA2020EPJ}. 
Most notably, the inherent short duration of the positron beam (inferred to be of the order of $\lesssim$ 50 fs, corresponding to $\sigma_z\lesssim 14$ {\textmu}m) is naturally suited for injection in a positron-accelerating wakefield structure without the need for complex beam manipulation, as also confirmed by the proof-of-principle simulations reported in this article.

\begin{table}[t!]
\begin{tabular}{|l|c|c|c|c|c|}
\hline 
 & Measured & Inferred & \cite{Muggli2008PRL} & \cite{Corde2015N} & \cite{Gessner2016NC}  \\
\hline 
E (GeV) & $ 0.6 $ & 0.6 & 28.5 & 20.3 & 20.3 \\
$\sigma_x$ ({\textmu}m) & 100 & 2.7 & 25 & $<100$  & 50 \\
$\sigma_z$ ({\textmu}m) & - & $\lesssim14$ & 730  & 30 - 50 & 35 \\
$\epsilon$ (nm)  & 480 & 15   & $14 \times 3$ & $5 \times 1$ & 7\\
$\bar{\epsilon}$ ({\textmu}m)
 & 560 & 18 & $390 \times 80$ & $200 \times 50$ & 300\\
\hline 
\end{tabular}
\caption{Summary of measured and inferred positron parameters, compared with sources used for previously reported proof-of-principle positron wakefield acceleration experiments.}
\label{tbl:beamComparison}
\end{table}


In conclusion, direct and comprehensive spatial and spectral characterisation of GeV-scale laser-driven positron beams is reported. 
Experimental results show that minimising the free propagation of the primary electron beam to the converter results in the production of GeV-scale positron beams with micron-scale source size and normalised emittance, using a 100 TW-class laser system. The beam is also shown to be of sufficient quality to undergo energy selection, with beamlets containing $\approx10^5$ positrons in 5\% bandwidths around \unit[500]{MeV} being isolated, and of sufficient quality to be injected in a plasma wakefield accelerator.
These results demonstrate the possibility of experimentally studying laser-wakefield acceleration of positrons, a critical milestone towards the realisation of the next-generation of plasma-based particle accelerators and colliders. For instance, an experimental platform of this kind can be implemented in future laser facilities with dual beam capability \cite{Danson2019HPLSE} or in beam-driven wakefield facilities with laser capability (e.g. FLASHForward \cite{Aschikhin2016NIMA} and SPARC\_LAB \cite{Ferrario2013NIMA}) to study beam-driven methods, without the need for an emittance damping storage ring.

In terms of future developments, increasing the energy of the primary electron beam is expected to further improve the positron beam characteristics. 
Single stage LWFA has been demonstrated beyond \unit[5]{GeV} \cite{Gonsalves2019PRL}, which along with a minimised drift distance for the LWFA electrons is expected to readily provide, for instance, nm-scale geometric emittances and a normalised emittance of \unit[10]{{\textmu}m} at \unit[3]{GeV} \cite{Sarri2022PPCF,Alejo2019SR}.
As a final remark, we note that different mechanisms for the laser-driven generation of positron beams based on the Breit-Wheeler pair production process have been numerically proposed (see, e.g., \cite{Ribeyre2017,Li2020}). 
However, these mechanisms require next-generation multi-PW laser facilities and generally result in positron beams of different characteristics. Studies on direct irradiation of solid foils with the next generation of ultra-high intensity laser systems has also been numerically reported \cite{Song2022}.

\noindent\textbf{Methods}\\
\noindent\textbf{Two-screen electron spectrometer:}
two scintillator screens were placed in the electron beam after the magnetic dipole, to detect electrons with kinetic energy in the range of \unit[$200\geq \gamma m_e c^2\geq  2500$]{MeV} and propagation angles relative to the laser propagation axis of \unit[$-15 \geq \theta_y \geq 15$]{mrad}. The 3D 3-vector field distribution of the dipole magnets was experimentally mapped and found in agreement with numerical simulations of the magnet setup using RADIA \cite{Elleaume1998IEEE}. The screen dispersion functions were calculated by numerically solving the particle trajectories using the Boris-pusher and recording the particle position on each screen as a function of the initial particle 3-momentum.
This was used to produce a look-up table which gave the particle energy as a function of its position on each screen and its initial propagation angle $\theta_y$.
The electron beam spectrum was determined by finding the coefficients of a third-order polynomial function $\theta_y(\gamma)$ that minimised the mean squared difference between the retrieved angularly integrated electron spectra from each screen.
Charge calibration of the electron spectrometers were performed by measuring electron spectra on an absolutely calibrated image plate placed in front of the LANEX screen and comparing to the images recorded on the CCD over the same shots. The image plate used was BAS-TR2040, with a sensitivity of 1 PSL per 350 electrons.\\


\noindent\textbf{Source size, divergence, and emittance retrieval:} The beam profile after the beam aperture was modelled as an azimuthally symmetric clipped Gaussian distribution, such that only particles with $x_{i,\mathrm{ap}}^2+y_{i,\mathrm{ap}}^2 \leq R_{\mathrm{ap}}^2$ were transmitted, where $x_{i,\mathrm{ap}}$ and $y_{i,\mathrm{ap}}$ are the transverse spatial coordinates of the $i^{th}$ particle at the aperture plane $z_{\mathrm{ap}}$, with aperture radius $R_{\mathrm{ap}}$.
This profile was dispersed according to the individual particle energies onto the spectrometer screen.
The particle distribution $S'_y(x)$ was measured at the detector plane $z_{\mathrm{det}}$ where $x$ and $y$ are transverse coordinates perpendicular and parallel to the dispersion plane of the spectrometer respectively.
Due to the combination of energy spread and divergence, the profile $S'_y(x)$ is due to particles over a range of different energies where their initial propagation angle $\theta_{i,y}$ and energy $E_{i}$ result in the particle intersecting the detector plane at the position $y$.
With the assumption that the spectrum $N(E)$ is slowly varying, then each slice measurement $S'_y(x)$ represents the integral of the beam profile over $y$, i.e.,
\begin{align}
    S'_y(x) &= 2 A_{y,0} \int_0^{\sqrt{R^2-x^2}} \exp\left[{-\frac{x^2 + y^2}{2\sigma_x^2}}\right]  \mathrm{d}y \nonumber \\
    S'_y(x) &= \sqrt{2 \pi} A_{y,0}  \sigma_x \erf\left(\sqrt{\frac{R^2-x^2}{2\sigma_x^2}} \,\right) \exp\left[{-\frac{x^2}{2\sigma_x^2}}\right] \label{eqn:beam_profile}
\end{align}
where $A_{y,0}$ is the amplitude of the particle distribution, $R=R_{\mathrm{ap}} z_{\mathrm{det}}/z_{\mathrm{ap}}$ is the projected size of the aperture at the detector plane and $x \leq R$.
The functional form of equation \ref{eqn:beam_profile} was used to fit the amplitude of the modulated signal $S_y(x)$ when retrieving the apertured beam properties as described below.

Several steps were followed to extract the particle emittance from the spectrometer signals.
Firstly, a variable threshold filter was used to remove hard-hits caused by stray photons hitting the CCD directly.
Secondly, the defocusing of the effect of the magnetic dipole fringe fields was removed by re-scaling the measured signal in non-dispersion direction such that the spatial frequency of the grid pattern was made constant for all energies.
Vertical slices were then taken through the resultant image, averaging over \unit[4]{mm} in the dispersion direction to produce the signal modulation $S_y(x)$ as a function of $x$ at a given $y$ position.
The scattered particles from the grid formed a smooth background on the detector which was removed by fitting a Gaussian to the values at the minima of the observed modulations. 
The envelope of the signal was similarly found by fitting the beam profile function (equation \ref{eqn:beam_profile}) to the signal maxima.
The RMS width of the fitted envelope was then divided by the source-to-screen distance to obtain the beam divergence $\sigma_{\theta}$.

An ideal zero source size beam would produce a sharp step-function within the bounds of the scattering signal and the beam envelope, with the spatial period of the magnified grid size.
Blurring of this pattern was observed due to contributions of the finite spatial resolution of the diagnostic (\unit[215]{{\textmu}m}) and the source size $\sigma_x$ of the beam, which was found by iterative minimisation of the mean squared error between the measured signal and the calculated signal for a given source size.
The geometric emittance was calculated as the product of the measured divergence and source size, i.e. $\epsilon = \sigma_x \sigma_{\theta}$.

In order to benchmark the retrieval process, synthetic data was created by numerically propagating results from a FLUKA simulation and removing particles that would hit the solid bars of the emittance measurement grid.
The dispersion of the magnet was added by shifting the particles transversely according to their energy using the same dispersion function as for the experimental spectrometer.
To create the modulated signal $S_x(y)$ for a given energy band, the particles are selected according to their position on the spectrometer.
Due to the significant beam divergence, there is some trajectory crossing such that some particles of different energies are selected, and some of the correct energy are omitted. 
The synthetic signals were analysed with the same procedure as for the experimental data and compared to the values directly calculated from the particle distributions. 
The retrieved beam properties closely agree with the directly computed values for the apertured beam, verifying the analysis procedure.\\

\noindent\textbf{Monte-Carlo simulations:} simulations of the bremsstrahlung induced pair-production process were performed using the particle physics Monte-Carlo code FLUKA with the EM-cascade defaults.
$10^6$ primary electrons for each converter thickness were simulated and the resultant particle number was then scaled up by 8738 to match the higher charge of the experimental LWFA electron beam.
A lead converter of variable thickness $L$ was placed in the path of the electrons, and the momenta and position of all electrons, positrons and photons were recorded as they exited the rear surface of the converter.
In order to simulate the effects of the finite divergence of the electron beam, each particle was assigned random angular shifts ($\Delta x'_i$ and $\Delta y'_i$) from the probability density function $f(x') = f_0[(x'/\theta_w)^2+1)]^{-2}$, which was seen to approximate the experimentally measured transverse profile of the primary electron beam with \unit[$\theta_w=2.9\pm0.3$]{mrad} ($f_0$ is the normalisation constant).
The particle transverse momenta and positions were then altered according to these shifts and using the experimental drift length between the LWFA exit and the converter rear face of \unit[50]{mm}.  
The transverse particle positions were also modified according to the expected LWFA electron source size of \unit[1]{{\textmu}m}, although this contribution was negligible.
Each particle was shifted 10 times from the value taken from the FLUKA simulation, with the shifted particle properties recorded each iteration to produce a final list with 10 times the number of particles as were produced by the FLUKA simulations.
Particle distribution properties were then calculated at the longitudinal plane for which the correlation term $\langle x x' \rangle$ was minimised. 
\\

\noindent\textbf{FBPIC simulations:}
The FBPIC Particle-In-Cell code \cite{Lehe2016CPC} was used to run simulations of injection and acceleration of the output positrons from FLUKA Monte-Carlo simulations in a laser-driven wakefield. 
As FBPIC employs a cylindrical grid with azimuthal decomposition, two cylindrical modes were used to capture the physics of the wakefield formation and acceleration. 
The simulation was performed in the Lorentz boosted frame with a Lorentz factor of $\gamma= 10$. 
The wavelength of the laser was $\lambda_L = 1000\ \unit{nm}$, with a normalized vector potential of $a_0 = 1.5$. 
The background plasma profile consisted of a 0.5 mm linear up-ramp starting at the right-edge of the initial simulation window from vacuum to a plasma density of \unit[$n_{e0} = 2\times 10^{17}$]{cm$^{-3}$} which was then constant for 100 mm. 
The plasma terminates with a 0.5 mm long down-ramp of back to vacuum. The particles per cell set for each coordinate direction were 2 along $\hat{z}$, 2 along $\hat{r}$, and 6 along $\hat{\theta}$. 
The simulation box lengths were \unit[$z_{length} = \left( \frac{6 c}{\omega_{p0}} + 2\lambda_{p0}\right) \approx 220.6$]{{\textmu}m} and \unit[$r_{length} = 2\times r_{max} = 2\times \frac{15 c}{\omega_{p0}} \approx 2\times 178.2$]{{\textmu}m} for $\hat{z}$ and $\hat{r}$ respectively, where $\omega_{p0} = \sqrt{\frac{e^2 n_{e0}}{\epsilon_0 m_e}}$ is the plasma frequency of the peak initial density in the accelerator, and $\lambda_{p0} = \frac{2\pi c}{\omega_{p0}}$ is the plasma wavelength. 
The number of cells in each direction were for $\hat{z}$ $N_z = \lfloor\frac{z_{length}}{\Delta z}\rfloor = 1764$ where $\Delta z = \lambda_L / 8$, and for $\hat{r}$ $N_r =\lfloor\frac{r_{max}}{\Delta r}\rfloor = 89,$ where $\Delta r = 2\times \lambda_L$. 
The time step size of the simulation was set to be \unit[$\Delta t = r_{max}/(2\times \gamma \times N_r)/ c \approx 0.334$]{fs}.
The positrons simulated in figure \ref{fig:PositronsAccelResults} were imported into the FBPIC simulation from the output files given a random longitudinal spatial spread of \unit[10]{{\textmu}m} to match the inferred produced beam duration from this experimental campaign. \\

\noindent\textbf{Authors Contribution}
G.S. devised and proposed the experiment, which was numerically modelled and designed by T.A., L.C., J.C. and M.J.V.S.. The experiment was carried out by N.C., E.E.L., A.F.A., M.D.B., L.C., H.A., B.K., P.P.R., and D.R.S., under the overall coordination of M.J.V.S., C.C., and G.S. The data analysis was predominantly performed by M.J.V.S and the particle-in-cell simulations were carried out by J.C. and Y.M., under the supervision of A.G.R.T. The manuscript was written by G.S., M.J.V.S, and J.C. with input from S.P.D.M., Z.N., and A.G.R.T.\\

\noindent\textbf{Acknowledgments} 
The authors wish to acknowledge support from EPSRC (EP/V044397/1, EP/N027175/1, EP/V049577/1), STFC (ST/V001639/1), US NSF (grant \#2108075) and from the staff of the Central Laser Facility.

\end{document}